
\input harvmac

\def \om {\omega}

\def\const {{\rm const}}
\def \s {\sigma}

\def \p {\phi}
\def \ha {\half}
\def \ov {\over}

\def \four{{\textstyle {1\ov 4}}}
\def \eit{{\textstyle {1\ov 8}}}
\def \twe{{\textstyle {1\ov 12}}}

\def \a {\alpha}
\def \lr { \lref}
\def\ep{\epsilon}

\def \r {\rho}
\def\const {{\rm const}} \def\m{\mu}\def\n {\nu}\def\l
{\lambda}

\def \b {\beta}

 \def \sm {$\s$-model\ }
\def   \td {\tilde }
\def \k {\kappa}
\def \lr { \lref}

\gdef \jnl#1, #2, #3, 1#4#5#6{ { #1~}{ #2} (1#4#5#6) #3}

\def\np {  Nucl. Phys. }
\def \pl { Phys. Lett. }

\def \prl { Phys. Rev. Lett. }
\def \pr  { Phys. Rev. }

\def \ijmp { Int. J. Mod. Phys. }

\baselineskip8pt
\Title{\vbox
{\baselineskip 6pt{\hbox{  }}{\hbox
{Imperial/TP/95-96/16 }}{\hbox{hep-th/9512081}} {\hbox{
   }}} }
{\vbox{\centerline {Heterotic -- type I superstring duality}
 \centerline {and  low-energy effective actions}
 }}

\vskip -20 true pt






\medskip
\centerline{   A.A. Tseytlin\footnote{$^{\star}$}{\baselineskip8pt
e-mail address: tseytlin@ic.ac.uk}\footnote{$^{\dagger}$}{\baselineskip8pt
On leave  from Lebedev  Physics
Institute, Moscow.} }

\smallskip\smallskip
\centerline {\it  Theoretical Physics Group, Blackett Laboratory,}
\smallskip

\centerline {\it  Imperial College,  London SW7 2BZ, U.K. }
\bigskip\bigskip
\centerline {\bf Abstract}
\medskip
\baselineskip10pt
\noindent
\medskip
We compare
order $R^4$ terms in the 10-dimensional  effective actions of
 $SO(32)$ heterotic
and type I superstrings
from the point of view
of    duality between the two theories.
Some of these terms  do not receive higher-loop
corrections being related by supersymmetry
to `anomaly-cancelling' terms which depend
 on the antisymmetric 2-tensor.
At the same time, the consistency of
duality relation implies
that  the   `tree-level'  $R^4$ super-invariant
 (the one which  has  $\zeta(3)$-coefficient
in the sphere part of the action)
should  appear also at   higher orders
of loop expansion,
 i.e.  should be multiplied by a
non-trivial function of the  dilaton.

\Date {December 1995}

\noblackbox
\baselineskip 14pt plus 2pt minus 2pt
\def \tr {{ \rm tr  }}
\def \k {\kappa}

\lr \met {R.R. Metsaev and A.A. Tseytlin, \np B298 (1988) 109.}

\lr \chap {A.H. Chamseddine, \np B185 (1981) 403;
E. Bergshoeff, M. de Roo, B. de Wit and P. van Nieuwenhuizen, \np B195 (1982)
97;
G.F. Chaplin and N.S. Manton, \pl B120 (1983) 105.}
\lr \witten {E. Witten, \np B443 (1995) 85, hep-th/9503124.
}

\lr \dabgib {A. Dabholkar, G.W. Gibbons, J. Harvey and F. Ruiz Ruiz,  \np
B340 (1990) 33;
A. Dabholkar and  J. Harvey,  \prl
63 (1989) 478.
}

\lr\dab{A. Dabholkar, \pl B357 (1995) 307,
hep-th/9506160.}
\lr \hul{C.M. Hull, \pl B357 (1995) 545, hep-th/9506194.}

\lr \ftt { E.S. Fradkin  and A.A. Tseytlin, \np B227 (1983) 252.  }
\lr \ler { W. Lerche, \np B308 (1988) 102.}
\lr\lerr{W. Lerche, B.E.W. Nilsson and  A.N. Schellekens, \np B289 (1987)
609;
W. Lerche, B.E.W. Nilsson, A.N. Schellekens and N.P. Warner, \np B299 (1988)
91.}

\lr\gross{ Y. Cai  and C. Nunez, \np B287 (1987) 279; Y. Kikuchi  and C.
Marzban, \pr D35 (1987) 1400.}
\lr\grosss{ D.J.  Gross and J.H. Sloan, \np B291 (1987) 41. }
\lr\elli { J. Ellis, P. Jetzer and L. Mizrachi, \np B303 (1988) 1. }
\lr\ell { J. Ellis and L. Mizrachi, \np B302 (1988) 65. }
\lr\eli { J. Ellis, P. Jetzer and L. Mizrachi, \pl B196 (1987) 492. }

\lr \abe { M. Abe, H. Kubota and N. Sakai, \pl B200 (1988) 461;
 \np B306 (1988) 405. }
\lr \gs{ M.B. Green  and J.H. Schwarz, \pl B149 (1984) 117;
\np B255 (1985) 93. }
\lr \gsh{ M.B. Green  and J.H. Schwarz,
\pl B151 (1985) 21. }

\lr \gro{D.J. Gross  and E. Witten, \np B277 (1986) 1.}
\lr \tse { A.A. Tseytlin, \np B276 (1986) 391. }
\lr \callan {C.G. Callan, C. Lovelace, C.R. Nappi and S.A. Yost, \pl B206
(1988) 41;
\np B308 (1988) 221.}
\lr \elw {U. Ellwanger, J. Fuchs and M.G. Schmidt, \np B314 (1989) 175.}
\lr \green {M.B. Green, J.H.  Schwarz and E.  Witten, {\it Superstring Theory}
(Cambridge U.P., 1988).}
\lr \abo{ A. Abouelsaood, C. Callan, C. Nappi and S. Yost, \np { B280 }
 (1987) 599.}
\lr \aob{ E. Bergshoeff, E. Sezgin, C.N. Pope and P.K. Townsend, \pl B188
(1987) 70. }
\lr \frt{ E.S. Fradkin and A.A. Tseytlin, \pl {
B163  }
(1985) 123.}

\lr \abb {N. Sakai and M. Abe, Progr. Theor. Phys. 80 (1988) 162.}
\lr \grom{D.J. Gross, J.A. Harvey, E. Martinec and R. Rohm, \np B256 (1985)
253; B267 (1986) 75.}
\lr \mrt {R.R. Metsaev, M.A. Rahmanov and A.A. Tseytlin, \pl B193 (1987) 207;
A.A. Tseytlin, \pl B202 (1988) 81.}

\lr \bos{J.A. Shapiro and C.B. Thorn, \pr D36 (1987) 432;
J. Dai and J. Polchinski, \pl B220 (1989) 387.}

\lr \cai {  J. Polchinski and Y. Cai, \np B296 (1988) 91.}

\lr \green {M.B. Green, J.H.  Schwarz and E.  Witten, {\it Superstring Theory}
(Cambridge U.P., 1988).}

\lr\gibbb{G.W. Gibbons and D.A. Rasheed, ``Electric-magnetic duality rotations
in non-linear electrodynamics", DAMTP-R-95-46,
  hep-th/9506035;
``SL(2,R) invariance of non-linear electrodynamics coupled to an axion and a
dilaton", DAMTP-R-95-48,
hep-th/9509141.  }
\lr \pow{J. Polchinski and E. Witten, ``Evidence for heterotic -- type I
string duality", IASSNS-HEP-95-81, NSF-ITP-95-135, hep-th/9510169.}

\lr \hot {C.M. Hull and P.K.  Townsend, \np B438 (1995) 109,  hep-th/9410167.}

\lr \vaw { C. Vafa and E. Witten, \np B447 (1995) 261.}
\lr \ant{
 I. Antoniadis, E. Gava, K.S. Narain and T.R. Taylor, \np B455 (1995) 109,
hep-th/9507115.}
\lr \elis{ J. Ellis and L. Mizrachi,   \np B327 (1989) 595;
R. Kallosh and A. Morozov, \pl B207 (1988) 164. }
\lr \ber{ M. Bershadsky, S. Cecotti, H. Ooguri and C. Vafa, Commun. Math. Phys.
165 (1994) 31.}
\lr \aan{
I.Antoniadis, E. Gava, K.S. Narain and T.R. Taylor, \np B413 (1994) 162.
}
\lr \wii {E. Witten, private communication.}

\lr \ttt{  A.A. Tseytlin, ``On SO(32) heterotic - type I superstring duality in
ten dimensions",
 Imperial/TP/95-96/6, hep-th/9510173.}
\lr \tsss{  A.A. Tseytlin, \ijmp A5 (1990) 589.}
\lr \mts{R.R. Metsaev and A.A. Tseytlin, \pl B185 (1987) 52.}

\lr \nils{
  R.E.  Kallosh,
Phys. Scr. T15 (1987) 118;
B.E.W. Nilsson and A.K. Tollsten, \pl B181 (1986) 63. }

\lr\berg{ E. Bergshoeff and M. de Roo, \np B328 (1989) 439.   }
\lr\roo {M. de Roo, H. Suelmann and A. Wiedemann, \pl B280 (1992) 39; \np B405
(1993) 326.}
\lr \sue{  H. Suelmann, ``Supersymmetry and string effective actions", Ph.D.
thesis, Groningen, 1994.   }

\lr \yas {O. Yasuda, \pl B218 (1989) 455. }
\lr \miz { L. Mizrachi, \np B338 (1990) 209.}
\lr \yass {O. Yasuda, \pl B215 (1988) 306.  }
\lr \ieng{A. Morozov, \pl B209 (1988) 473;  R. Iengo  and C.-J. Zhu, \pl B212
(1988) 313.}
\lr \horw{P. Horava and E. Witten, ``Heterotic and type I string dynamics from
eleven dimensions", IASSNS-HEP-95-86,  hep-th/9510209.}

\lr \metts{E. Bergshoeff, M. Rakowski and E. Sezgin, \pl B185 (1987) 371;
R.R. Metsaev and  M.A. Rahmanov, \pl B193 (1987) 202;
R.R. Metsaev, M.A. Rahmanov and A.A. Tseytlin, \pl B193 (1987) 207.}
\lr \zan{M.T.  Grisaru, A.E.M. van de Ven and D.  Zanon,
\np B277 (1986) 388, 409; M.T.  Grisaru and D.  Zanon,
\pl B177 (1986) 347; M.D. Freeman, C.N. Pope, M.F. Sohnius and K.S. Stelle, \pl
B178 (1986) 199.   }
\lr \ft { E.S. Fradkin  and A.A. Tseytlin,
\pl B158 (1985) 316;
\pl B160 (1985) 69.  }
\lr \sak {N. Sakai and Y. Tanii, \np B287 (1987) 457.}

\newsec{Introduction}
It was suggested  \witten\ that  $SO(32)$ heterotic and type I string theories
in ten dimensions
are  dual to each other in the sense that a strong-coupling
region  of one of the theories  can be  described
by dynamics of solitonic states which is equivalent
to  weak-coupling  dynamics of elementary states of the other.
Further arguments in
favour of this duality were given in \refs{\dab,\hul,\pow,\horw}.
The $D=10$ supersymmetry dictates that the leading-order terms in the
two effective actions are  related  by a  field redefinition.
Somewhat surprisingly, this redefinition involves changing the sign of the
dilaton, i.e. it
inverts the  string coupling \witten\   (we shall use primes to indicate
the fields of type I theory)
\eqn\red{ {G'}_{\m\n} = e^{-\p} G_{\m\n}\ , \ \ \
\p'=-\p  \ , \ \ \ \   B'_{\m\n}=B_{\m\n}\ , \ \ A'_\m = A_\m \ .  }
To   understand better how this  this  duality acts
at the  string-theory level
it would be interesting to compare  higher-derivative  terms in the two
low-energy effective actions.\foot{Similar  tests of
heterotic--type II string duality \refs{\hot,\witten}
in six and four dimensions were done  in \refs{\vaw }
and \refs{\ant}.}

To be able to test  duality  using only perturbative string-theory
results one is
to consider the  terms in the effective actions  with  coefficients
  that have  simple polynomial   dependence on string coupling,
i.e. receive
 contributions only from one or few  particular
orders  of string loop expansion \wii. Then
  \red\  maps
them into terms with  similar perturbative coefficients.\foot{It should be
noted that
one may ignore possible $\a'$-dependent modifications of the duality
transformation rule \red. Indeed, the effective actions are
in any case defined modulo local field redefinitions \refs{\tse,\gro}, so it
makes sense to compare only `irreducible' terms  with coefficients which are
not changed under local covariant redefinitions of `massless' fields.
Such  are the terms which will be discussed below.}

 There are, indeed, examples of terms in the string effective actions which are
either  not renormalised by
 string higher-loop corrections  (like second-derivative terms and
`anomaly-cancelling' terms,  see, e.g., \refs{\green, \elis,\lerr,\yas})
or receive contributions only from specific
orders of string perturbative expansion  (see \refs{\ber,\aan}).
In \ttt\  we considered a
 term quartic  in  the gauge field strength $\tr F^4$ which,
in  heterotic string  theory,   is absent at the
tree level \refs{\gross,\grosss} but appears at the one-loop  level
 \refs{\elli,\abe,\ler}. We  have
 argued  that it does  not receive corrections
from higher loops since $D=10$ supersymmetry relates \refs{\roo,\sue} it
to the `anomaly-cancelling' term $B\tr F^4$  \gs\
which is not renormalised at higher orders \refs{\miz,\yas}
(see also below).
The duality \red\  maps this term  into
the tree-level (disc) $\tr F^4$ term of  the type I effective action
\refs{\tse,\gro}.
In  the finite $SO(32)$ type I theory \gsh\ the $\tr F^4$ term does not receive
any loop corrections and its coefficient is exactly the same \ttt\ as that of
the one-loop $\tr F^4$ term in the heterotic theory.

As was noted in \ttt,
 it seems  likely
that $D=10$  superstring effective actions contain   infinite series of {\it
local} { terms}\foot{In addition to  local terms,
the  massless superstring effective action contains also
non-local terms which are non-analytic in momenta.
We define the  string effective action
as the one the tree-level amplitudes of which reproduce the full loop-corrected
string amplitudes for massless states. The non-analyticity of the low-energy
expansion is due to loops of  massless
string states  which
must necessarily be included in order to have a well-defined  (finite,
anomaly-free)
effective action \refs{\tsss}.}
 that
receive contributions
only from one particular    order in  string loop  expansion.
This is  necessary for  consistency of the duality conjecture and is
probably  related  to the special property of $D=10$ supersymmetry
that certain  super-invariants
  of  given  dimension  may have only  specific dilaton dependence
\refs{\roo,\wii}.
Indeed, the  dilaton plays  a   special role
in $D=10$  supergravity,  being in the same multiplet with graviton.

Here we shall  try to  elucidate further
 the interplay  between duality,
supersymmetry and the structure of loop expansion in heterotic and type I
theories
by extending
the discussion  in \ttt\
to the curvature-dependent $R^4$-terms.
We shall see that the  relation between  the $R^4$ corrections
is more  complicated  than  between the  $F^4$  terms,
reflecting the existence of several super-invariants containing $R^4$
contractions  \refs{\berg,\roo,\sue}.

\newsec{Duality and effective actions}

The general structure of local terms depending on the curvature and gauge field
strength
in the heterotic  and type I
string effective actions is
\eqn\acot{ S_{het}=
 \int d^{10} x \sqrt G\  \sum^\infty_{n=1} \big[g_n (\p) R^n
+  s_n(\p)   R^{n-2}F^2 +  ... +  f_n (\p) F^n  +... \big] ,    }
$$ S_{typeI}=
 \int d^{10} x \sqrt {{G'}}\  \sum^\infty_{n=1} \big[g'_n (\p') R'^n+
s'_n(\p')   R'^{n-2}F'^2
+ ...  +  f_n (\p') F'^n  + ...\big] ,    $$
where $R^n = (R^{\cdot} _{\dots })^n G^{-n}$ stands for various possible
invariants  with   $n$ factors of  the curvature.
$F^n$ may also involve different group trace structures which we shall not
distinguish at the moment.
 We assume that the inverse string tensions $\a'$ and $\a'_I$
 are absorbed into
the metrics $G_{\m\n}$ and ${G'}_{\m\n}$  so that all tensors have geometrical
dimension $[T_{\m_1...\m_n}]= cm^{-n}$.

In order for the two actions to be related by the duality transformation
  \red\  it should be true that
\eqn\duud{
g'_n(\p') = e^{(n-5)\p'} g_n(-\p') \ , \ \ \
 s'_n(\p') =   e^{(n-5)\p'} f_n(-\p') \ ,  \ \  .... \ . }
 Our aim is to check  whether  such relations are  consistent with
the structure of string perturbative expansions in the two
theories.

Since we known how to compute the above coupling functions
  only  using
   weak-coupling expansions
in each theory
\eqn\hh{ g_n (\p) = b_0^{(n)} e^{-2\p} + b_1^{(n)} + b_2^{(n)} e^{2\p}  + ... +
b_m^{(n)} e^{2(m-1)\p} + ... \ ,  \ \ \ e^\p \ll  1\ , }
\eqn\heh{ g'_n(\p')  = { b'}_0^{(n)} e^{-\p'} + { b'}_1^{(n)} + { b'}_2^{(n)}
e^{\p'}  + ... +{ b'}_m^{(n)} e^{(m-1)\p'} + ... \ , \ \ \  e^{\p'} \ll  1\ ,}
to be able to check  the duality relations
\duud\ one should  consider only  special invariants
which have coupling functions
containing only one or few terms in the perturbative series.
Indeed, in general,
the left and the right
sides of the  formal relation for $g'_n(\p')$ in \duud\
\eqn\hwh{   { b'}_0^{(n)} e^{-\p'} + { b'}_1^{(n)} + { b'}_2^{(n)} e^{\p'}  +
... +{ b'}_m^{(n)} e^{(m-1)\p'} + ... }
$$
= b_0^{(n)} e^{(n-3)\p'} + b_1^{(n)}e^{(n-5)\p'} + b_2^{(n)} e^{(n-7)\p'}  +
... + b_m^{(n)} e^{ (n-3 - 2m)\p'} + ... \  ,
$$
are defined in the different regions of the  coupling space,  $e^{\p'} \ll  1$
and $e^{\p'} \gg  1$,  respectively.

Let us start with  the $R,R^2,R^3$ terms.  Since  the 3-graviton amplitude
in  the heterotic string theory does not receive
string loop corrections   \refs{\grom,\gross,\grosss, \elis,\elli}\foot{The
proof that there is no  Gauss-Bonnet-type $R^3$ term involves analysis
of the 4-point graviton amplitude \refs{\mts,\grosss}. As usual, we
 ignore terms which can be eliminated by local field redefinitions.}
\eqn\corr{ g_1(\p) = d_1 e^{-2\p}  , \ \  g_2(\p) = d_2 e^{-2\p} , \ \
 g_3(\p) = d_3 e^{-2\p} , \ \ \
d_2= {\textstyle {1\ov 8}} d_1, \ \ d_3=0 \ . }
Then \duud,\hwh\  would be satisfied provided
\eqn\corr{ g'_1(\p') = d_1 e^{-2\p'}  , \ \ \ \
 g'_2(\p') = d_2 e^{-\p'} , \ \ \ \ g'_3(\p') = d_3 \ .
 }
These relations   can be  understood as  being simply a consequence
of $D=10$ supersymmetry: the dilaton dependence
of the supergravity term is fixed uniquely (up to a field redefinition)
as is the structure of the $R^2$  super-invariant; also,
there are  no  super-invariants  containing
$R^3$ (see \refs{\nils,\berg} and references there).
According to \berg,  the supersymmetric action
which starts with  the  $N=1, D=10$ supergravity + Yang-Mills  terms
with  $H_{\m\n\l}$
modified by the Lorentz  Chern-Simons term $\omega_3$
is given by an  infinite series of terms  (we use heterotic frame)
\eqn\acty{ S_{het}=
- \eit
 \int d^{10} x \sqrt G\    e^{-2\p} \big[R  + 4 (\del \p)^2 - \twe \hat
H^2_{\m\n\l} }
$$
- \  \ha   T
+  k_2(3 T_{[\m\n\l\r]}^2 + T_{\m\n}^2) + ... + O(T^n) +...
  \big], $$
$$T_{\m\n\l\r} \equiv  k_1 \tr (F_{\m\n} F_{\l\r})  +  k_2  \tr (R_{\m\n}
R_{\l\r})
\ , \ \ \ \  T_{\m\r} \equiv  T_{\m\n\l\r}G^{\n\l} \ , \ \ \ T= T_{\m\n}
G^{\m\n}
\ ,
$$
where $R_{\m\n} $ stands for $R^{ab}_{\ \ \m\n}(\om_-)$.
The connection $\om_-$ has
torsion proportional to  $\hat H_{\m\n\l}$ which is given by   $\  \hat H=dB +
k_1 \omega_3(A) + k_2 \omega_3(\om_-)$. The coefficients
$k_1$ and $k_2$  are fixed by the condition of anomaly cancellation  at the
level of
 low-energy field theory and thus do not contain any
particular  `stringy' information.
They must have the required form in any
 theory which reduces to supergravity + Yang-Mills at low energies, is
supersymmetric and
 anomaly-free.
Since the heterotic and type I effective actions start with the same
supergravity + Yang-Mills
actions related by  the field redefinition \red\
we conclude that
the consistency of  duality
at this level
is  automatic.

One may,  of course,  check the coefficient in
 the expression for $g_2'$
by directly computing the sum (finite in $SO(32)$  theory \gsh)  of the
3-graviton amplitudes on  the
disc and the  projective plane.\foot{In addition  to the $O(e^{2n\p})$
contributions from the orientable diagrams
of the type II theory (sphere, torus, etc.) the closed string type I scattering
amplitudes receive contributions from
 extra diagrams (with topology of a sphere with
 disc and crosscap insertions) which should be added  together
in order to realise the projection onto  type I intermediate states \green.}
To confirm that $d_3=0$, i.e. $g_3'=0$ in type I theory
 one  is to   show that the
 $R^3$-contribution
which could  come
 from the sum of the annulus, M\"obius strip  and Klein bottle diagrams
 does not appear at all.

 Another  consistency check is provided by an observation that
 the sphere part of the type I action
(which is the same as in  type II theory with  the NS-NS antisymmetric tensor
field  set equal to zero) does not contain
$R^2$ and $R^3$ terms. The terms
 $e^{-2\p'} (d_2' R'^2 + d_3' R'^3)$
in the type I action  would be  related by
the duality  \red,\duud\
to
 the  terms  $ d_2'e^{-\p} R^2  + d_3' R^3$ in the heterotic action
which are certainly  absent in the
 heterotic string perturbative expansion.
As mentioned above, the  presence of $R^2$ and the absence of  $R^3$-term  in
the two effective actions
is, in fact, dictated by the  10-dimensional supersymmetry.

\newsec{$R^4,R^2F^2,F^4$ terms in the $SO(32)$ heterotic string action:
 super-invariants and non-renormalisation }
Let us now consider   $R^4$, $R^2F^2$ and  $F^4$  terms.
We shall  start by   summarising  the known results
for the structure of  tree-level  and 1-loop corrections
to the effective action of the  heterotic string and
then explain how these results can be understood
systematically  in terms
of possible $D=10$  super-invariants constructed in \refs{\roo,\sue}.

The relevant tree-level terms in the heterotic string action
can be written in the following symbolic form \refs{\gro,\gross,\grosss,\zan}
($\a'$ is absorbed into $G_{\m\n}$;the  trace in the fundamental representation
of $SO(32)$
and $\tr R^2 \equiv R^{ab}_{\ \ \m\n} R^{ba\m\n}$)
$$ S^{(0)}_{het}= - \eit
 \int d^{10} x \sqrt G\    e^{-2\p} \big[R  + 4 (\del \p)^2 -  \twe  \hat
H^2_{\m\n\l}
+  \eit  (\tr F^2  -   \tr R^2) $$
\eqn\actu{  + \  b_1 t_8  (\tr F^2  -   \tr R^2)^2
+ b_2 (t_8 t_8 R^4 -  {\textstyle {1\ov 8}} \ep_{10} \ep_{10} R^4)
 \big]\  , }
$$ b_1= - {1\ov  2^{8 }}, \ \ \ \ \  \ \  b_2 = {\zeta(3)\ov 3\cdot 2^{9}} \ .
$$
Here $t_8$
is the 10-dimensional extension of the 8-dimensional
light-cone gauge `zero-mode' tensor \green\
(with $\epsilon^{\m_1...\m_8}$-term omitted)
 built out of $G^{\m\n}$. For example,
\eqn\ffff{ t_8  F^4\equiv
 t^{\m_1\n_1 ... \m_4\n_4} F_{\m_1\n_1} F_{\m_2\n_2} F_{\m_3\n_3}
F_{\m_4\n_4}      }
$$  = \ 16 F^{\m\n}F_{\r\n}F_{\m\l}F^{\r\l} +
8 F^{\m\n}F_{\r\n}F^{\r\l} F_{\m\l} $$
 $$
- \  4 F^{\m\n}F_{\m\n}F^{\r\l}F_{\r\l}
- 2 F^{\m\n}F^{\r\l}F_{\m\n}F_{\r\l} \ ,  $$
and
\eqn\rrr{ t_8 t_8 R^4 \equiv  t^{\m_1\n_1 ... \m_4\n_4} t_{\m'_1\n'_1 ...
\m'_4\n'_4}
R_{\m_1\n_1}^{\m'_1\n'_1}...
R_{\m_4\n_4}^{\m'_4\n'_4}\ . }
We  shall  use   $\ep_{10}$  to indicate the totally antisymmetric {\it tensor}
$\ep^{\m_1...\m_{10}}$, e.g.,
\eqn\eppi{
\ep_{10} \ep_{10} R^4\equiv
 \ep^{\a \b \m_1\n_1 ... \m_4\n_4} \ep_{\a \b \m'_1\n'_1 ... \m'_4\n'_4}
R_{\m_1\n_1}^{\m'_1\n'_1}...
R_{\m_4\n_4}^{\m'_4\n'_4} \ .  }
The  invariant \eppi\ is the  Gauss-Bonnet
density in 8-dimensions whose presence cannot be detected
 from the  calculation   of the  4-point scattering amplitude
but  can be  fixed by comparison with the  \sm $\b$-function
 \refs{\zan}.

The above terms are  consistent
with the $D=10$ supersymmetry.
The structures  containing $\tr F^2  -   \tr R^2$
are `anomaly-related' terms
which appear in the super-extension of the supergravity + super Yang-Mills
with $H$  modified by the Lorentz Chern-Simons term \berg, i.e. \acty\
with $k_1=-k_2=\four$.
The  combination
\eqn\sus{ J_0= t_8 t_8 R^4 - {\textstyle {1\ov 8}} \ep_{10} \ep_{10} R^4 \ , }
is  the bosonic part of one of possible
  super-invariants containing $R^4$-terms
\refs{\nils,\roo,\sue}.

The  one-loop terms in the effective action
which are local and have explicitly computable coefficients
are related to the special  amplitudes on the torus
which have `nearly holomorphic' integrands
 \lerr\ so that they
receive contributions only from a  boundary (Im$\tau\to \infty$) of the moduli
space.
They can be  reconstructed from string scattering amplitudes
\refs{\lerr,\ell,\elli,\abe} or  computed directly \ler\
using the partition function representation for the string
effective action \ft.
The parity-odd `anomaly-cancelling'  terms in the  heterotic string
theory
have the following structure
\refs{\lerr}\foot{We
 quote  the 1-loop coefficients relative to the tree-level term \actu.
In the notation  we are using here
$\k=  2 \a'^2 g, \  g^2_{10} = 2\k^2/\a', \  g= e^{\p_0},$
i.e. $S^{(0)}_{het}=
 \int d^{10} x \sqrt {\td G} \    e^{-2{\td \p}}
\big(- {1\ov 2\k^2} {\td R} + ... )$.}
\eqn\atu{ S^{(1)}_{het}=
 \int d^{10} x \sqrt G\  \big( - {\textstyle {1\ov 16}}  \ep_{10} B  X_8 + ...
\big)  \ , }
where $\ep_{10}$ is multiplied by  $B_{\m\n}$
and   the  8-rank tensor $X_8$ which is related to the anomaly  12-form (and
elliptic genus $\cal A$)
by $I_{12} = [{\cal A}_{q^0}]_{12} = {1\ov 2\pi} (\tr F^2  -   \tr R^2) X_8$,
\eqn\xxx{
X_8 =
\b \big( 32 \tr F^4 - 4 \tr F^2 \tr R^2 + 4 \tr R^4  + \tr R^2 \tr R^2
\big) \ , \ \ \ \ \b \equiv -{ 1 \ov 3\cdot 2^{14} \pi^5} \ .   }
The parity-even  local one-loop terms  are given by
 \refs{\elli,\abe,\ler}
\eqn\atut{ S^{(1)}_{het}=
 \int d^{10} x \sqrt G\  \big( \four   t_8  X_8 + ... \big) \ , }
where    $t_8X_8\equiv t_{\m_1...\m_8} X_8^{\m_1...\m_8}$.
The similarity of the expressions \atu\  and  \atut\
suggests that they are   related by supersymmetry (which  was indeed
 anticipated in \ler).
A  heuristic argument explaining  the   connection between
  $t_8 X_8$ and $\ep_{10}BX_8$ terms is  the following.
The  8-dimensional light-cone gauge  `zero-mode' tensor  which  appears
 in the 4-point amplitude \green\
$(t^{\m_1...\m_8})_8 = t^{\m_1...\m_8} - \ha \ep^{\m_1...\m_8} $
 may be given  a  10-dimensional
generalisation:
\eqn\req{ \hat t^{\m_1...\m_8} \equiv  t^{\m_1...\m_8} - \four B_{\l\r}
\ep^{\l\r\m_1...\m_8}\ }
(assuming that in  the light-cone gauge $B_{uv}=1, \ F_{uv}=R_{uv}=0$).
 This is also consistent with the structure of the corresponding string
amplitudes
\refs{\elli,\ell}
 which, of course,  should satisfy the requirements of
linearised supersymmetry. Then the combinations $t_8 X_8$ and $\ep_{10}BX_8$
in \atu\  and  \atut\
  should be parts  of the $D=10$ super-invariant
\eqn\sws{ J_1\equiv   \hat t_8 X_8 = t_8 X_8 - \four \ep_{10} B X_8 \ ,
}
\eqn\ute{ S^{(1)}_{het}=
 \int d^{10} x \sqrt G\  \big(  \four J_1  + ... \big) \
 . }
Indeed,  $J_1$ is a combination of
 $R^4,F^4,R^2F^2$ type $D=10$ super-invariants recently constructed in
\refs{\roo,\sue}.
According to \refs{\roo,\sue}
the bosonic parts of a set of  6
independent super-invariants are given by
(an invariant action should start with $R + \tr F^2$-terms)
 \eqn\suss{ I_1= t_8 \tr F^4 - {\textstyle {1\ov 4}} \ep_{10} B \tr F^4 \ , \ \
\  I_2= t_8 \tr F^2 \tr F^2  - {\textstyle {1\ov 4}} \ep_{10} B \tr F^2 \tr F^2
\ ,}
\eqn\suso{ I_3= t_8 \tr R^4 - {\textstyle {1\ov 4}} \ep_{10} B \tr R^4 \ , \ \
\  I_4= t_8 \tr R^2 \tr R^2  - {\textstyle {1\ov 4}} \ep_{10} B \tr R^2 \tr R^2
\ ,
}
\eqn\sust{ I_5= t_8 \tr R^2\tr F^2  - {\textstyle {1\ov 4}} \ep_{10} B \tr
R^2\tr F^2 \ ,
}
which all contain parity-odd terms,  and also
by the parity-even  combination
$J_0$ in \sus. Note  also that
\eqn\rela{ t_8t_8 R^4 = 24 t_8 \tr R^4 - 6 t_8 \tr R^2 \tr R^2 \ . }
As a result, $J_1$
 \sws\ which appears in the 1-loop heterotic string action \ute\ is
a linear
combination
of the  super-invariants (see \xxx)
\eqn\suuu{ J_1 = \b (32 I_1  + 4 I_3  + I_4 - 4 I_5) \ . }
One expects that the parity-odd `anomaly-cancelling' terms
$\ep_{10}B \tr F^4$, etc., should appear only at the  1-loop order,  since,
in particular,
there should be no higher-loop
contributions to the anomalies \yass.
The absence of corrections to these terms
was  shown
   directly at the level of higher-loop heterotic string
amplitudes  \refs{\miz,\yas}.
The relation between the string coupling  and the dilaton $\p$,
and  also the gauge nature of the 2-form  field $B_{\m\n}$,
suggest  another simple  explanation for that.
If one would get, e.g.,  $f(g) \ep_{10}B \tr F^4$ with
 $\  f(g)= a_1 + a_2g^2 + ..., $
this would imply the presence of the  $e^{2n\p} \ep_{10}B \tr F^4$
terms,  which, however,
are not   consistent with the gauge invariance $B\to B + d \lambda$
unless $n=0$.

Combining the non-renormalisation of the  `anomaly-cancelling' terms
with their relation to   $R^4,F^4,R^2F^2$-terms in \suss,\suso,\sust\ by
$D=10$ supersymmetry
we are led to the important conclusion that
the coefficients of the latter terms
 also
{\it do not receive two and higher loop
corrections}.\foot{A close connection  between the calculation
of the anomaly index and the one-loop
$O(R^4,R^2F^2,F^4)$  term
represented as a  torus partition function in a background
was suggested as an indication that this term should  not receive higher
heterotic string loop corrections \ler.
It should be emphasised that the condition of preservation of supersymmetry is
crucial for this non-renormalisation.
One may also  argue for non-renormalisation of, e.g.,  $F^4$ term
 by modifying
the  proof   \yas\  of  the absence of renormalisation of $\ep_{10}BF^4$ term.
The   $g$-loop  parity-conserving 4-vector ($V = \zeta \cdot ( \del x + ik \psi
\psi ) e^{ikx}$)
 amplitude
in RNS approach  has
   $2g-2$ supermoduli integrals and thus contains
  $2g-2$ supercurrent ($T_f = \psi \del x + .... $)
insertions.
Since we need only 4 powers of momenta $k$  to get $F^4$-structure
and at the same
time  are  to saturate the  integral
 over 10 fermionic zero modes,  the  two
(additional to 8)  $\psi$-factors   should come from the
supercurrent insertions.
Then  the remaining free  $\del x $ from $T_f$
should be contracted with  $e^{ikx} $ giving   extra power of momenta.
This  implies that though there may be
higher-derivative corrections,  $F^4$  will not be renormalised.}
This is an interesting    new  example of a  $D=10$ `non-renormalisation
theorem'  which  applies to terms  originating  from certain
 {\it 4-point}   string amplitudes.\foot{Since the field strengths contain
non-linear commutator terms  and the curvature terms
expanded near flat space contain terms of all orders in $h_{\m\n}$,
the gauge invariance implies
similar non-renormalisation of certain
low-momentum terms in  infinite sets of  vector
and graviton  amplitudes.}

This non-renormalisation  allows one  to  extend
the duality relation between  the leading terms in
the type I and heterotic string
effective actions  to  those higher-derivative terms.

Note, however,  that we are unable to make
a similar  non-renormalisation claim
for the parity-even `tree-level'
 super-invariant  $J_0=t_8 t_8 R^4
- {\textstyle {1\ov 8}} \ep_{10} \ep_{10} R^4$  in \sus.  As we
shall see below, this  invariant  must, in fact,
{\it  receive  higher loop  corrections}
in order  to avoid contradiction with  the duality conjecture.

The $ t_8  t_8  R^4$   term  is present    both  in  the
 tree \refs{\zan,\gro}  {\it and } the  1-loop \refs{\sak,\abe}
 parts of   the type II
superstring effective action. The $D=10$ supersymmetry implies that  it  should
actually appear in  the   combination
$J_0$  \sus, i.e.
\eqn\ttu{S_{typeII}=
 \int d^{10} x \sqrt G\  \big[ g(\p) J_0 + ...  \big]  \ , }
$$
 g(\p)=  c_0 e^{-2\p}   +   c_1 + ...\ , \ \ \  \
c_0 = - {1\ov 8}  b_2\ ,
\ \ \      c_1  =- {1 \ov 3 \cdot 2^{18} \pi^5 }\  ,  $$
where $b_2$ is the same as in \actu.
In contrast to the heterotic string case, here
the  needed $R^4$ kinematic structure is already
produced by the  light cone Green-Schwarz fermionic
zero-mode  contribution so that its coefficient  is simply given by
 the volume
of the moduli space.
As a result,
there are no 1-loop contributions proportional to $I_3$ and $I_4$ in \suso\
(in agreement with the absence of the
 `anomaly-cancelling' terms).\foot{Note  also that  while the tree-level terms
in the heterotic \actu\
 and type II \ttu\
actions  coincide  for the special background   $R=F$,
this does not apply to the 1-loop corrections (cf. \sus,\xxx,\rela).
Though the two world-sheet actions become formally equivalent, the
structure of the two 1-loop path  integrals   remains
different.}

\newsec {Comparison with  type I theory}
One may argue that the
comparison of the coefficients
of the above terms  ($R^4$, etc.)   in the heterotic and type I actions
does not  represent a  non-trivial check of the duality as such:
the values of these coefficients are determined just
by the requirement of anomaly cancellation
in low-energy field theory.
Indeed, the coefficients of the `anomaly-cancelling' terms
 ($\ep_{10}BR^4$, etc.)
  are  directly related to the
anomalous contributions in massless fermionic  amplitudes
of low-energy field theory and thus are   fixed  once one demands
the anomaly cancellation via Green-Schwarz mechanism \gs.
Since the supersymmetry relates them to the coefficients of the $R^4$, etc.
terms, the latter coefficients are also uniquely determined by the low-energy
field theory.\foot{It is
thus should  not be  surprising   that
  computed  directly from  string theory  they are given
  by  the  contribution of the `infra-red' boundary of the moduli space.}
Since both type I and heterotic string theories   are consistent
(supersymmetric, anomaly-free)  extensions
of the same low-energy field theory, the values of the
coefficients in the two theories which are uniquely determined by
the limiting field theory  {\it must} be the same.

Still,  given that the structures of the  loop expansion
in the two  string theories are very different,
it is  instructive to see
how  the duality transformation
relates the terms which appear at different loop orders
 in a way consistent with  $D=10$ supersymmetry.
A test of the duality is
actually the equality of these coefficients {\it combined}
with their non-renormalisation by higher-loop corrections.

Let us start with the tree-level terms in the heterotic string  \actu\
and discuss their type I  images  under the duality transformation \duud.
Since  the `anomaly-related'
terms like $ \sqrt G e^{-2\p} t_8  (\tr F^2  -   \tr R^2)^2 $ in \actu\
are parts of the supersymmetric completion
of the  low-energy supergravity action (with Chern-Simons modified $H$)
these terms should not be
renormalised by higher loop corrections for the same reason as
the leading-order supergravity terms.
Then  according to \duud,\hwh\   the   counterpart of the above term
under the duality transformation \red\
is  $\sqrt{{G'}} e^{\p'}t_8  (\tr {F'}^2  -   \tr {R'}^2)^2 $. This
term should originate from  the combination of
type I  diagrams with the
 Euler number  $-1$ (sphere with 3 holes,  etc).
These diagrams are  much more complicated than the sphere one which
led  to \actu\ in the heterotic string. This  illustrates
the  non-triviality of the duality relation between
the two theories.

The same reasoning  does not automatically apply  to $J_0$ in \sus,\actu\
since   it is not clear how  one    could
 argue that it does not receive corrections
from higher loops. This term thus needs a special  discussion
 (see below).

Since the  1-loop torus terms in  \atu,\suuu\
do not appear at the tree level and  two and higher loop levels,
the  corresponding function $g(\p)$ in  \hh\ is constant.
Then under the  duality \duud,\hwh\
they are mapped into `tree-level' disc terms
$e^{-\p'}\tr F^4 , e^{-\p'}R^4, e^{-\p'}R^2 \tr F^2)$.
Thus the absence of the
 $\tr F^2 \tr F^2 $ term in the torus correction in
$SO(32)$ heterotic string  is  crucial for  the consistency of the duality:
 in type I  theory  such  double-trace
term could not originate from the disc
 diagram which has only one boundary.
As was  pointed out  in \ttt,  the coefficient
of the  disc $t_8\tr F^4$ term in type I theory \refs{\tse,\gro}
is indeed the same as the 1-loop coefficient
 of this term (see \atut,\xxx) the heterotic string \refs{\elli,\abe}.

The `anomaly-cancelling' terms in \atu\
do not depend on the dilaton  and constant part of the metric
and thus do not change  their form (for $\p=\const$)
under the duality transformation \red.
Their coefficients  (determined, as mentioned above,
 uniquely  by the same low-energy field theory)
 must  indeed  agree
 in the two string theories  even though
they originate  from two different string diagrams  -- torus in the  heterotic
string \lerr\
and disc (with  $B_{\m\n}$ R-R vertex operator
insertion adding extra power of $e^\p$) in type I theory \callan.
Since the field redefinition \red\
implies the corresponding change in the supersymmetry transformation
laws,  the type I theory  form  of the super-invariants \suss\
is
\eqn\ssf{ I_1=  e^{-\p'} t'_8 \tr {F'}^4 -
{\textstyle {1\ov 4}} \ep'_{10} B \tr {F'}^4  , \ \ \  \
I_3=  e^{-\p'} t'_8 \tr {R'}^4 -
{\textstyle {1\ov 4}} \ep'_{10} B \tr {R'}^4 \ , ...\ ,    }
where the  prime is used to indicate that the
corresponding  tensors  are constructed using ${G'}_{\m\n}$.
As in the heterotic  case,
the  non-renormalisation of `anomaly-cancelling' terms
 combined with supersymmetry  implies  the
non-renormalisation of the
$\tr F^4, R^4, \tr F^2 R^2$-terms appearing in \suss,\suso,\sust\
in type I theory.\foot{A direct argument  relating
 the non-renormalisation of $\tr F^4$ to finiteness of type I theory
was sketched  in \ttt.}

 Let us now discuss  whether there
are  other $R^4$, etc. terms in the type I action  in addition to
 the ones which are dual images  of the
terms in the heterotic action considered  above.
The   `anomaly-related' terms  and the terms
which are  connected by
supersymmetry  to the  unique `anomaly-cancelling' terms   are also unique; the
duality  transformation \red\  just relates the corresponding super-invariants.
This does not, however, apply to the
$R^4$-terms which form the super-invariant $J_0$ \sus\
appearing  in the tree-level heterotic string action \actu.
Its direct type I counterpart under \red\
\eqn\cou{  \sqrt G e^{-2\p} J_0  = \sqrt {G'} e^{\p'} J'_0, \ \  \ \ \
J_0'= t'_8 t'_8 {R'}^4 - {\textstyle {1\ov 8}} \ep'_{10} \ep'_{10} {R'}^4 \ , }
is the bosonic part of only  one of the
super-invariants of that  structure which  are present in the effective
action of type I theory.
Indeed,
 the type I theory
contains $J_0$ terms coming from the sphere and the torus since the
contributions of these diagrams are the same  as in the type II theory.
The latter terms were
 already given in \ttu.
The corresponding  type I  contributions  are obtained
 by adding primes on $G$ and $\p$, i.e.\foot{The value of the coefficient $c_1$
may, in principle, be different from its value in the type II theory:
in addition to the torus, the  Euler number zero diagrams in the type I theory
include
also the  Klein bottle, the annulus and the  M\"obius strip which may also
produce contributions  proportional to $J_0$.}
\eqn\ttur{S_{typeI}=
 \int d^{10} x \sqrt {G'}\  \big[ g'(\p')  J'_0   + ... \big]  \ , \ \ \   \ \
g'(\p')=   c_0  e^{-2\p'}  +   c'_1 + ...  \ . }
As it  is  clear  from \cou,\ttur,
the $J_0$-term in type I theory is likely to receive corrections
from all orders of perturbation theory,
$g'(\p')=   c_0  e^{-2\p'}  +   c'_1 +  c_2'e^{\p'} + ...\ $.
This then avoids   contradiction with duality making it simply non-applicable
term-by-term in  the two weak-coupling  expansions.
If  the duality \red\  would  apply to the terms in \ttur,
they  would be mapped  into
\eqn\tyr{S_{typeI} (G,\p)=
 \int d^{10} x \sqrt G\  \big[ ( c_0  e^{\p}+   c'_1 e^{-\p} + ...) J_0 (G)  +
... \big]  \ .   }
Such terms  are certainly absent
in the heterotic string perturbative expansion being of
odd power in the  string coupling constant.

The absence of contradiction with duality implies that
$J_0$ must  receive higher-loop corrections also in the heterotic  theory
(otherwise  duality would map the tree-level heterotic string $J_0$ term
in \actu\  into the unique type I term  \cou).
This indicates that there should exist a super-extension
of the term $g(\p) J_0$ with an arbitrary  dilaton  function.
Such a possibility does  not seem to be rulled out by  a
scarce information
available about the structure of   `anomaly-unrelated'
super-invariants in $D=10$ supergravity.
In particular, the (on-shell or only linear)
superspace constructions of $R^4$ superinvariants \refs{\nils}
are not sufficient in order to fix uniquely the structure of
possible  dilaton prefactor.
Though the fact that the dilaton is in the same multiplet with graviton seems
to suggest that that the form of the dilaton dependence
 can not be arbitrary,
this  expectation seems to be  in conflict with the above discussion
of the  $g(\p) J_0$-terms in the type I superstring.

While one is thus  unable to
check duality by comparing $J_0$-terms
in the two theories,   one can use the duality conjecture to {\it predict}
 the strong-coupling  behaviour  of the corresponding $g(\p)$
function in each theory: it should be  the same as  its weak coupling
behaviour in the dual theory.  Thus, for example,
in the heterotic string theory (see \duud,\tyr)
\eqn\ggg{ g(\p)_{\p \to -\infty }  = c_0 e^{-2\p} + ... , \ \ \ \ \ \
g(\p)_{\p \to \infty }  = c_0 e^{\p}  + c'_1 e^{-\p} ...\   . }

\newsec{Concluding remarks}
Similar conclusion about a  non-trivial modification to all loop orders
applies  to all terms  which appear in the low-energy expansion
of the sphere contribution to the  type I effective action.
Indeed,  the
 $R^{n} $ terms  there
  have the dilaton factor $g'_{n} =  b'_{n} e^{-2\p'}$. If they were not
renormalised,
the duality \red,\duud\   would  map them  into
$ b'_{n} e^{(n-3)\p} R^n$ terms in the heterotic action.
Since the supersymmetry seems to rule out higher-order terms with  odd  powers
of $R$
this means that  {\it all}  $R^{n}  $  ($n > 1$) terms in the sphere part
of the  type I action  would be  mapped into
 the heterotic string  terms  multiplied by
  {\it odd} powers  of   string coupling.
Such terms, though  presumably
consistent with  $D=10$
supersymmetry,
 cannot appear in  the perturbative  loop expansion of the heterotic string.
Thus
 all of them should  be `dressed' by  non-trivial functions of  dilaton.

At the same time, all $\tr F^n$ terms
in the disc  (`Born-Infeld') part of the type I action  should not receive
higher loop corrections \ttt. This implies, by duality, that the
$\tr F^n$ terms in the heterotic string action  should
appear only at specific  orders  of the loop
expansion,
i.e. should have the following exact  dilaton dependence:
  $a_n e^{(n-4)\p} \tr F^n$.
Like $a_4$, other $a_n$ coefficients probably be given by the contributions of
boundaries of  moduli spaces, in agreement with their
`tree-level' (disc) origin in the type I theory.
Each of these  terms should be consistent with
supersymmetry since
 the  $F^n$-cominations
which originate from  the expansion of the (abelian)
Born-Infeld action
are indeed bosonic parts of  (global $D=10$) super-invariants \metts.

Thus one expects that  $D=10$ superstring effective actions
should contain   local higher-derivative  terms with coefficients
which  receive contributions only from certain loop orders and
 only from boundaries  of moduli spaces
 so that they  are explicitly computable.
A low-dimensional  example of such `non-renormalisation' is provided by
  $N=2, D=4$ supersymmetric terms $\sim R^2 F^{2g-2}$
 in type II theory
compactified on a Calabi-Yau space
which  are generated  only  at $g$-loop  order  \refs{\ber,\aan}.
This follows from the $N=2$ supersymmetry  (the dilaton is a member of a
hypermultiplet)
and  is also
 related to the fact that the gravi-photon  $F$
is a R-R field
(so that its power  should be  correlated with the power of $e^\p$).

It is  likely that  analogous
 `non-renormalisation theorems'
 may apply  also directly  to type II theory  in 10 dimensions.
Given that    superspace non-renormalisation arguments
are  currently not  applicable in  the $D=10$ case,
one may try to use indirect arguments,
demanding consistency between string loop expansion and  duality
combined also with  some   information about $D=10$ supersymmetry.
In particular,  it would be
interesting  to  apply   self-duality  of the
type IIB theory  \refs{\hot, \witten}
to try to
determine the dependence of its effective action on the
 two antisymmetric 2-tensors
(NS-NS  and R-R ones)
which are interchanged under the  special
duality transformation $\p' = -\p,$ etc.
Since the dilaton  dependence of the  terms with the
R-R field  is likely to be  fixed,
this  may  also constrain possible   dependence
on the NS-NS field.

\bigskip
\noindent {\bf Acknowledgements}
\bigskip
I am grateful to  E. Bergshoeff,  J. Louis, M. de Roo,  K. Stelle
and  E. Witten
for important remarks and suggestions.
I would like to thank  the organizers of the
CERN workshop on duality
for hospitality  and invitation to present a talk.
I acknowledge also the support of PPARC,
ECC grant SC1$^*$-CT92-0789 and NATO grant CRG 940870.
\vfill\eject
\listrefs
\end

At the same time, there are certainly tree-level (sphere) terms
involving  powers of $dB_1$. By duality they are mapped into powers
of $dB_2$ with fixed dilaton dependence.
Indeed, under string-string duality of type IIB theory (...)
tree-level $e^{-2\p} H^{2m}_{\m\n\l}$ terms get transformed into
.....
This is consistent with expectation that world-sheet coupling of RR state
in Green-Schwarz approach  should be
$$ L = G_{mn}(x) \del_a x^m \del^b x^n
 +  \bar \theta \gamma_n  \rho^a  \rho^b \del_a \theta  \del_b x^n  $$
$$
+ ... +  e^\phi  \bar \theta \rho^a\gamma_m  \gamma^{p...q} \gamma_n
\rho^b\theta
   \del_a x^m \del_b x^n H_{p...q} + ... $$
 Thus each $H$-factor is accompanied
by $e^\p$ factor.
That means that the   $k$-loop  $O(H^m)$ contribution
will have the  dilaton  factor
$e^{[2(k-1) + m ]}\p$.